\documentclass[conference]{IEEEtran}
\IEEEoverridecommandlockouts
\usepackage{cite}
\usepackage{amsmath,amssymb,amsfonts}
\usepackage{graphicx}
\usepackage{textcomp}
\usepackage{xcolor}
\usepackage{blindtext}
\usepackage{lettrine}
\def\BibTeX{{\rm B\kern-.05em{\sc i\kern-.025em b}\kern-.08em
    T\kern-.1667em\lower.7ex\hbox{E}\kern-.125emX}}
\begin{document}

\title{Demonstration of microwave multiplexed readout of
DC biased superconducting nanowire detectors
}

\author{
A. K. Sinclair,
E. Schroeder,
D. Zhu,
M. Colangelo,
J. Glasby,
P. D. Mauskopf,
H. Mani,
and K. K. Berggren
}
\maketitle

\footnote{Support for this work was provided in part by the DARPA Defense Sciences Office, through the DETECT program.

A. K. Sinclair, E. Schroeder, J. Glasby, P. D. Mauskopf, and H. Mani are with Arizona State University, Tempe, AZ 85281 USA (e-mail:adrian.sinclair@asu.edu).

D. Zhu, M. Colangelo, and K.K. Berggren are with the Department of Electrical Engineering and Computer Science, Massachusetts Institute of Technology, Cambridge, MA ZIP USA.}
\begin{abstract}
Superconducting nanowires are widely used as sensitive single photon detectors with wide spectral coverage and high timing resolution. We describe a demonstration of an array of DC biased superconducting nanowire single photon detectors read out with a microwave multiplexing circuit. In this design, each individual nanowire is part of a resonant LC circuit where the inductance is dominated by the kinetic inductance of the nanowire. The circuit also contains two parallel plate capacitors, one of them is in parallel with the inductor and the other is coupled to a microwave transmission line which carries the signals to a cryogenic low noise amplifier. All of the nanowires are connected via resistors to a single DC bias line that enables the nanowires to be current biased close to their critical current. When a photon hits a nanowire it creates a normal hot spot which produces a voltage pulse across the LC circuit. This pulse rings down at the resonant frequency of the LC circuit over a time period that is fixed by the quality factor. We present measurements of an array of these devices and an evaluation of their performance in terms of frequency and time response.
\end{abstract}


\section{INTRODUCTION}
\lettrine{S}uperconducting nanowire single photon detectors (SNSPDs) are currently the leading technology in single photon counting and timing in the ultra-violet, optical, and near infrared. Recent experiments have demonstrated a variety of approaches to implementing arrays with these devices, including an array of independently biased and electrically read-out, a cross-bar array with resistors that inhibit sneak current \cite{b1}, a delay-line based approach in which delay elements are used to separate detector signals in time \cite{b2}.
An overview of other readout architectures for SNSPDs is given in \cite{b3}.
These arrays are potentially important for applications in communications, spectroscopy, and of course imaging. 
However, all of these schemes require significant development of external readout hardware. 
Meanwhile, for a number of years frequency multiplexed readout electronics have been developed for kinetic inductance detectors \cite{b4}\cite{b5}\cite{b6}\cite{b7}\cite{b8}\cite{b9}. 
A few of these readouts are actively deployed as facility instruments \cite{b9}\cite{b8} and more are planned to be \cite{b10}.
An SNSPD-based array that utilizes frequency multiplexing could thus take advantage of this substantial infrastructure.

Another group has also recognized this potential benefit and has made considerable progress \cite{b11} \cite{b12}. 
Our approach differs in the method of biasing the nanowires. 
We apply a DC bias in parallel and \cite{b11} AC biases the detectors from the microwave feedline. 
There are advantages and disadvantages to each method which are discussed in section \ref{comparison}.  

In this work we have developed a microwave frequency multiplexed array of DC biased SNSPDs. The operation, simulation, fabrication, assembly, and measurements of this device are described in the following sections.

\section{DEVICE THEORY OF OPERATION}\label{theory}
As is typical with SNSPDs, we DC bias the nanowire just below its critical current. 
When a photon is absorbed with an energy greater than the superconducting band gap \cite{b13} it breaks apart Cooper pairs, generates quasiparticles, and excites phonons. 
This region quickly becomes normal and begins to grow due to the continued Joule heating from the DC bias.
When the normal region expands across the narrow width of the nanowire a voltage pulse is generated due to the high normal state resistivity.
After some time the quasiparticles recombine to again form Cooper pairs and the hotspot disappears.
The nanowire alone can be modeled as an LR circuit and its impulse response described by an exponential decay with time constant $\tau = L/R$.
Where L is the total inductance of the nanowire including geometrical and kinetic and R is the resistance of the hotspot and transmission line or amplifier input impedance.
A thorough investigation of this electro-thermal model is given in \cite{b14}.

In order to multiplex many detectors onto one microwave line each is placed into a resonant circuit. 
With this configuration the impulse response is instead described by an RLC circuit. 
The transient response contains an exponential and sinusoidal component.
In this case the quality factor of the resonator will determine the exponential decay time, $\tau_{ring} = 2Q/\omega_0$. The resonant frequency follows, $\omega_0 = 1/\sqrt{LC}$.

\begin{figure}[!t]
\centering
\includegraphics[width=2.5in]{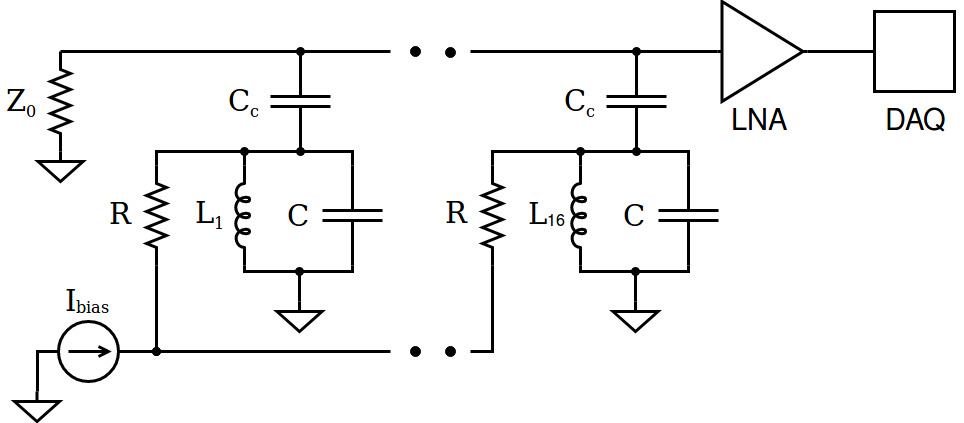}
\caption{Equivalent electrical circuit of the microwave multiplexed SNSPD array. Each nanowire acts as an inductive element $L_i$ in parallel with a capacitor C to form a resonant circuit. Tuning of the resonant frequency is achieved by varying the inductance of each nanowire i.e. $L_i \neq L_j$. Each resonator is connected to a single DC bias $I_{bias}$ through a resistor R. A coupling capacitor $C_{c}$ couples the resonators to the same transmission line with characteristic impedance $Z_{0} = 50 \ \Omega$. Signals on the transmission are amplified by a low noise amplifier (LNA) before being passed to the data acquisition system (DAQ).}
\label{fig:circuit}
\end{figure}

An equivalent electrical circuit is shown in Fig. \ref{fig:circuit}.
Two capacitors are used for each detector, one in parallel with the nanowire and one which AC couples the circuit to a microwave transmission line.
Each nanowire has a resistor which connects it to the DC bias. The resistor increases the impedance along the DC path, which minimizes cross talk and current redistribution between the detectors.

As a proof of concept, simulations were performed with a superconducting nanowire SPICE model \cite{b15}.
The simulation consisted of two nanowires placed into resonant circuits which are both coupled to a single transmission line.
The impulse response initiated by a photon absorption event matched the expected RLC transient. Two port scattering parameters were also obtained from this simulation.
With the promising results we continued onto fabrication and assembly of the device.

\subsection{Comparison Between AC and DC Bias}\label{comparison}
While this work and \cite{b11} both utilize frequency multiplexing, the two differ in the method of biasing. The method of AC biasing and standard DC biasing have been compared in an earlier publication \cite{b16}. It was found that AC biasing reduces the system detection efficiency due to the time varying bias current \cite{b16}.  

In the DC approach, fabrication complexity increases as each detector requires a DC line in addition to the resonant circuit components. This approach also suffers from a increased heat load due to the bias resistors dissipation. In contrast, the AC approach allows detectors to be independently biased from the microwave transmission line according to their resonant frequency. 

Each approach faces a unique challenge knowing that the internal detection efficiency of an SNSPD is a function of bias current \cite{b17} \cite{b18}. The ability to supply a DC bias without limiting the efficiency will depend on the consistency of critical currents across the array and also on the ability to distribute the bias current. For the AC bias the consistency between critical currents is not as important because each is biased by a unique probe tone that corresponds to a specific frequency. AC biasing instead modulates the internal detection efficiency on a timescale proportional to the inverse of the resonant frequency.

Both methods require similar room temperature readout electronics to demultiplex and demodulate the detector responses. The AC bias requires an arbitrary waveform generator in addition to the demodulator. When scaling arrays in the AC biasing scheme the high crest factor of the arbitrary waveform would only increase the dynamic range requirements of the readout amplifier.

\section{FABRICATION AND ASSEMBLY}

16 rows of linear arrays of SNSPDs were fabricated on a 2 cm $\times$ 2 cm SiO2-on-Si substrate for use in this experimental device.
The specific fabrication process followed is similar to the one described in \cite{b19}.
Each array consists of 16 detectors with identical nanowire elements (100 nm wide meandered nanowire covering a 3 $\mu$µm $\times$ 3 $\mu$m with a 50$\%$ fill factor) which are connected to varying length inductors with designed inductances ranging from 110 nH to 520 nH.
This allowed for the fine tuning of the resonant frequency of each detector with identical surface mount capacitors.
We randomly pre-screened 134 devices in a cryogenic probe station and 130 of them showed consistent switching currents.

\begin{figure}[!t]
\centering
\includegraphics[width=2.5in]{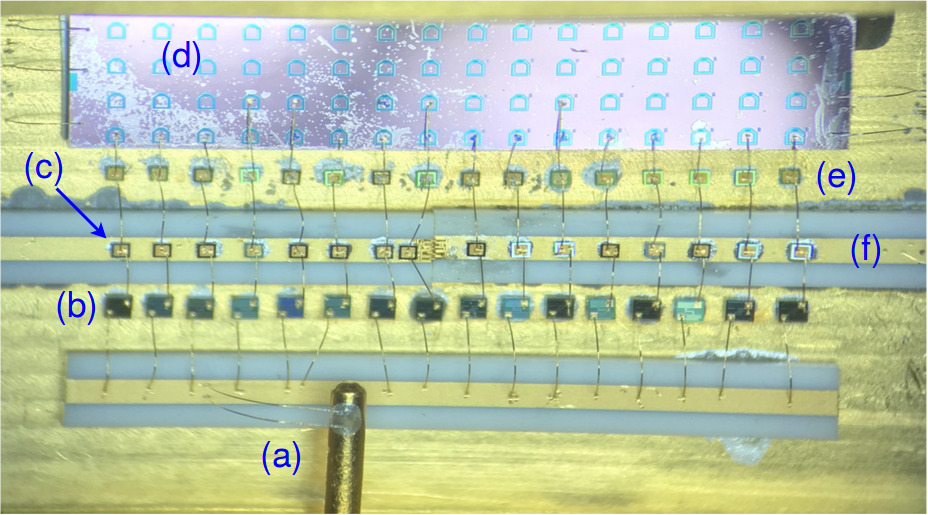}
\caption{Microscope image of the assembled 16 detector device. The DC bias pin (a) is connected to a transmission line which acts to supply the bias to all devices in parallel. Each resonator is connected to the DC bias through a resistor (b). Coupling the resonators to the transmission line (f) is a coupling capacitor (c). A diced portion of the substrate containing a linear array of nanowires (d) of which we use 16. Al wirebonds connect the capacitor (e) in parallel with the nanowire. The microwave transmission line (f) is connected to SMA ports on the outside of the package.}
\label{fig:device}
\end{figure}

A gold plated aluminum box with SMAs at either end houses the device. An image of the assembled device is given in Fig. \ref{fig:device}.
A 50 $\Omega$ transmission line traverses the length and acts as our microwave feedline for the resonators. 
Near the inductor array the transmission line has 1.8 pF surface mount capacitors for the coupling the resonator. 
In between the nanowire array and transmission line is another set of 6.8 pF capacitors in parallel with the nanowire and shunted to ground. 
A DC bias is supplied from the side of the package and is connected across a 2.5 k$\Omega$ resistor in series with the nanowire to ground. 
Conductive epoxy was used to secure the surface mount device components and nanowire array. Wirebonds connected to the nanowire array are Al and all other bonds are Au.

\section{EXPERIMENTAL SETUP}
The SNSPD array was placed inside a small light tight box, heat sunk to the 4 K stage of a cryostat cooled with a closed cycle refrigerator. 
A Keithley 2400-LV was used to DC bias the SNSPD array. 
A near infrared LED was positioned above the array within the box to illuminate the detectors. 
The output of the SNSPD array is amplified by a wideband cryogenic low noise amplifier (LNA). 
The cryogenic LNA has 30 dB of small signal gain over the 1 MHz - 2 GHz bandwidth and 5K of input referred noise. 
After cryogenic amplification the signal is then passed through a series of stainless steel coaxial SMA cables to a hermetically sealed feedthrough at 300 K. 
Additional amplification is achieved with a minicircuits ZKL-1R5+ (10 to 1500 MHz, 40 dB of gain). 
The output of this amplifier is sampled by a Tektronix TDS7401 10 GS/s Oscilloscope, Agilent E5072A VNA, and ROACH2 with MUSIC A/D and D/A board \cite{b4}.

\section{PRELIMINARY MEASUREMENTS}

Two devices were assembled from the linear nanowire arrays, one with a 2.5 k$\Omega$ bias resistor and the other with 50 k$\Omega$. 
For the 2.5 k$\Omega$ device we found that the frequency response nearly matched the LTSpice simulation at high frequencies and diverged at lower frequencies. 
In DC current bias mode many detectors pulses were observed on the oscilloscope but due to the low quality factor it was difficult to discern the frequency of each pulse. 
This motivated the assembly of another device with 50 k$\Omega$ bias resistors which would increase the quality factor and with it the ringdown time. 
With longer ringdowns the frequency determination would be easier at the cost of a lower maximum count rate.

After the 50 k$\Omega$ device was assembled the same tests were conducted. 
A measurement of $|S_{21}|$ showed only 3 resonances out of 16 at 84, 86, and 124 MHz.
\begin{figure}[!t]
\centering
\includegraphics[width=2.5in]{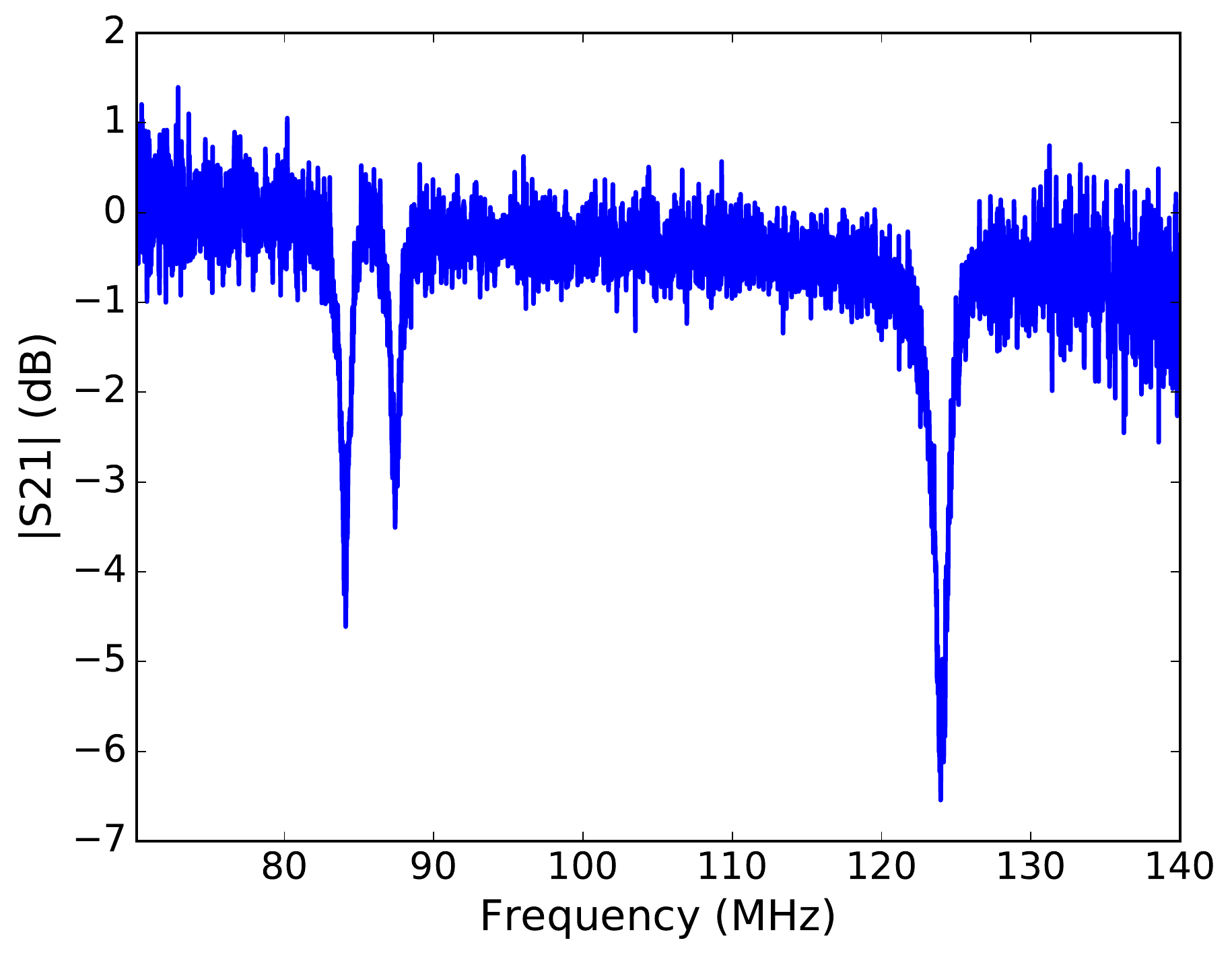}
\caption{Measured two port scattering parameter $|S_{21}|$ as a function of frequency. Three resonances were observed at 84, 86, and 124 MHz. These fall within the expected frequency range of 75 to 160 MHz.}
\label{fig:s21}
\end{figure}

It was at first suspected that the Al bonds were faulty or had broken due to thermal contraction. This turned out to not be the case with post cooldown measurements of the device. By measuring the total resistance as we removed the Al bonds to the nanowires one by one, we found only three were actually connected. This suggests that only three devices were initially connected before the cooldown.

For the existing resonances the Q’s were an order of magnitude higher than the previous device \cite{b20} approximately 139 , 147, and 142 respectively. 
This is also shown by the long ringdown of approximately 200 ns in the oscilloscope traces in DC current bias mode. The voltage noise rms was on order of 100 mV and this is over 10 times greater than the expected Johnson-Nyquist noise. In the spectrum of the oscilloscope traces a spike is present at 2.4 MHz. Radio frequency (RF) pickup is suspected, future measurements will include improved RF shielding.

\begin{figure}[!t]
\centering
\includegraphics[width=2.5in]{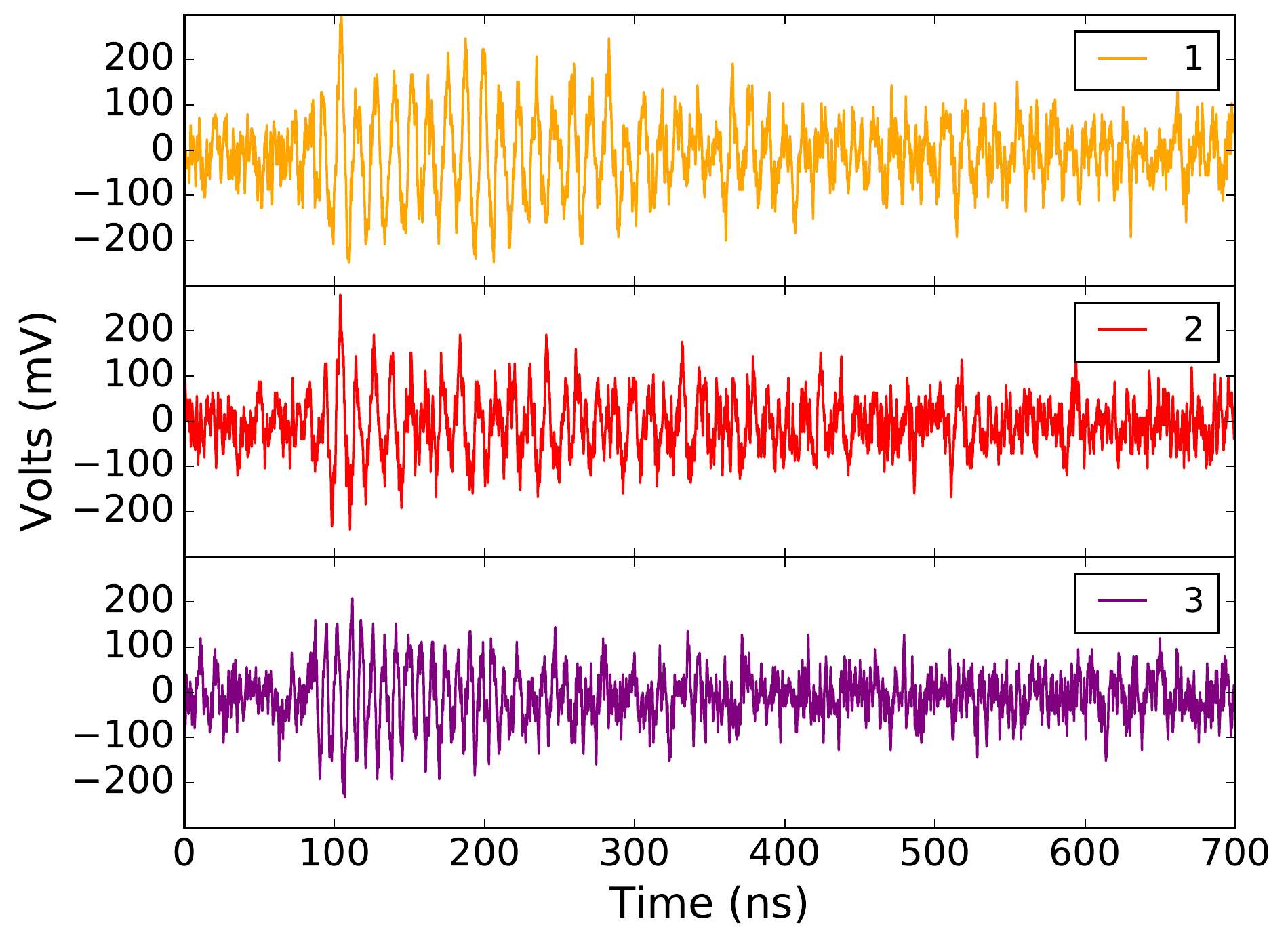}
\caption{Single shot oscilloscope traces of three distinct impulse responses. All three exhibit the characteristic rising edge, exponential decay, and sinusoidal ringdown.}
\label{fig:pulse_timestreams}
\end{figure}

To extract the frequency of the sinusoid from the pulses we Fourier transformed single shot oscilloscope traces.
There are three distinct peaks in the spectrum at 84, 86 and 124 MHz which are not surprisingly the same frequencies as our resonators.

\begin{figure}[!t]
\centering
\includegraphics[width=2.5in]{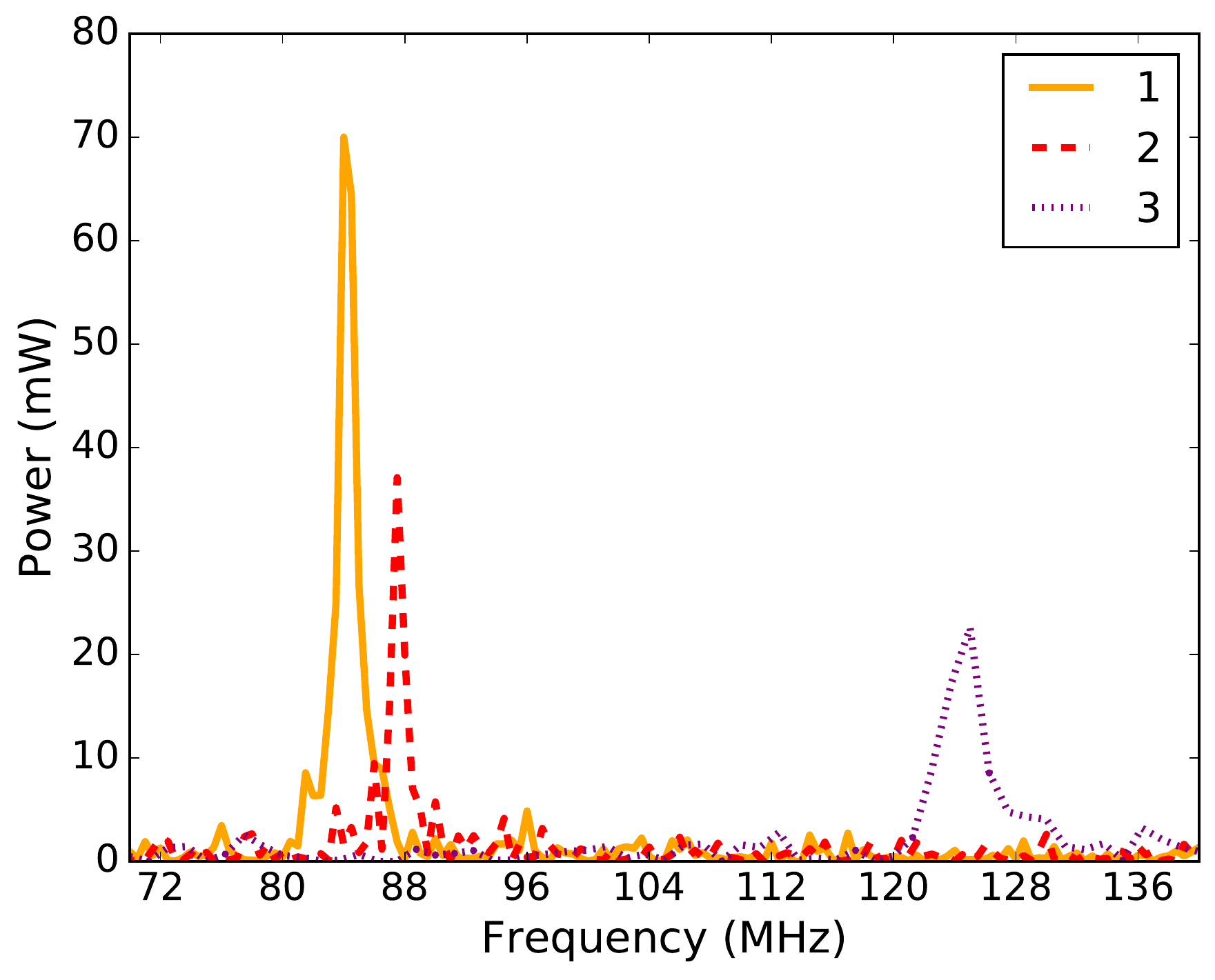}
\caption{Power spectrum of the three oscilloscope traces shown in Fig. \ref{fig:pulse_timestreams}. Each trace is independently Fourier transformed and over-plotted.}
\label{fig:power_spectrum}
\end{figure}

\begin{figure}[!t]
\centering
\includegraphics[width=2.5in]{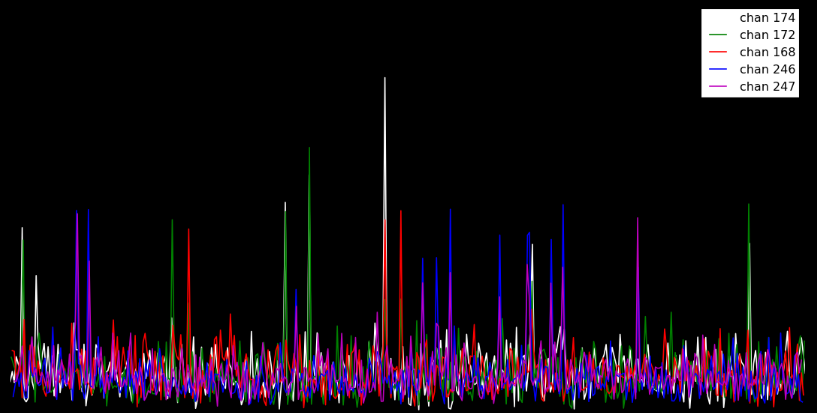}
\caption{A screenshot of 5 FPGA FFT bin timestreams over-plotted. Each channel corresponds to a different FFT bin and therefore different center frequency. The FPGA FFT bin width is 500 KHz. FFT bins were chosen that were close to the three resonant frequencies. In some cases if a resonance is near an FFT bin edge we also plot the adjacent bins. Each peak above the noise floor would be considered a detection event in that channel.}
\label{fig:screenshot}
\end{figure}

An additional measurement was performed with a custom FPGA based readout. 
This consisted of a ROACH2 with a MUSIC A/D and D/A board along with FPGA firmware that is intended to readout large arrays of kinetic inductance detectors for BLAST-TNG \cite{b6}.
We also used a slightly modified version of kidPy, a Python based terminal user interface program for operating the ROACH2 system with kinetic inductance detectors. 
After the signal is digitized at 512 MHz it passes through a polyphase filterbank and then though a 1024 point FFT. 
The firmware is parallel so that the FPGA clock rate is 256 MHz. 
This gives a spectral update rate of 500 KHz. 
Using the Python terminal interface a snapshot of the spectrum can be obtained on demand. As shown in Fig. \ref{fig:screenshot} repeatedly calling this function produces a spectrogram of the FFT bins chosen. 
As a preliminary readout design we show that this method can detect single photon pulses from the FFT bins that correspond to the resonant frequencies of the detectors.

\section{CONCLUSION}
We demonstrated a frequency multiplexed DC biased SNSPD array. A comparison was drawn between the two methods of biasing. 
We also have presented preliminary results of its operation with a modified kinetic inductance detector room temperature readout.
Future work will include developing FPGA firmware specific to frequency multiplexed SNSPD arrays with sub-nanosecond timing precision. We will improve RF shielding for the cryogenic setup and re-wirebond the 16 detector device.
There are also plans to fabricate an array of 100+ SNSPDs placed into resonant circuits with inter-digital capacitors. 500 MHz of instantaneous bandwidth could readout 100 detectors with a conservative spacing of 5 MHz between adjacent detectors. We are optimistic that the multiplexing factor will only improve given recent advances in kinetic inductance detector fabrication \cite{b21}.

\section*{Acknowledgment}
A. K. Sinclair thanks E. Lunde and C. H. Wheeler for help with the cryogenic system and R. Stephenson for informative firmware simulations.

\vspace{12pt}
\end{document}